\documentclass[twocolumn]{article}

\usepackage[utf8]{inputenc}
\usepackage[T1]{fontenc}
\usepackage{amsmath,amsfonts,amssymb}
\usepackage{algorithm}
\usepackage{algorithmic}
\usepackage{array}
\usepackage{booktabs}
\usepackage{tabularx}
\usepackage{makecell}
\usepackage{tikz}
\usetikzlibrary{positioning,calc,arrows.meta}
\usepackage{textcomp}
\usepackage{stfloats}
\usepackage{xurl}
\usepackage{cite}
\usepackage{graphicx}

\newcolumntype{Y}{>{\raggedright\arraybackslash}X}
\newcommand{\cmark}{\ensuremath{\checkmark}}
\newcommand{\source}{\mathcal{S}}
\newcommand{\event}{\mathcal{E}}
\newcommand{\timeline}{\mathcal{H}}

\begin{document}

\title{The Role of Vehicles in Digital Forensic Investigations: A Structured Synthesis of Digital Vehicle Forensic Characteristics}

\author{Kevin Mayer\vspace{.5em}\\\emph{Technische Hochschule Rosenheim}}

\onecolumn
\maketitle
\begin{abstract}
Modern vehicles are cyber-physical, networked systems that may contain valuable digital traces for accident reconstruction, crime investigation, warranty analysis, and cybersecurity incident response. However, digital vehicle forensics (DVF) remains less mature than computer, mobile, and cloud forensics because relevant data is distributed across in-vehicle components, mobile devices, manufacturer back ends, third-party services, and physical evidence. This article addresses this gap through a structured synthesis of academic literature, standards, and practitioner-oriented sources. First, we define DVF as the identification, preservation, acquisition, verification, interpretation, and reporting of vehicle-related digital evidence under safety, legal, privacy, and forensic-soundness constraints. Second, we formalize the DVF triage problem as the selection and correlation of evidence sources subject to volatility, accessibility, safety, integrity, and authorization constraints. Third, we explain how eight characteristics were derived from the literature and case material: multiple users, massively networked, cyber-physical system, dependencies between components, functional data, safety implications, accessibility, and limited abstraction. Finally, we add an adversarial perspective and a characteristic-driven triage procedure that helps investigators prioritize evidence sources while documenting assumptions, limitations, and failure cases. The resulting contribution is not an algorithmic performance claim; it is a reproducible conceptual framework for understanding, planning, and communicating DVF investigations.
\end{abstract}
\twocolumn

\section{Introduction}\label{sec:introduction}
Modern vehicles introduce a variety of digital services and features, such as smartphone integration, remote vehicle assistant applications, traffic-light communication, vehicle-to-everything communication, telematics, function-on-demand services, and manufacturer back-end connectivity. These services enlarge the attack surface and generate traces that may be relevant to cybersecurity, crime, warranty, insurance, and accident-related investigations. At the same time, investigators cannot assume that every involved vehicle has the capabilities of the newest vehicle generation. The German Federal Motor Transport Authority reported an average passenger-car age of approximately ten years in Germany in 2023 \cite{Statista2023}. Hence, digital vehicle forensic practice must work across legacy and modern vehicles, manufacturers, and heterogeneous evidence sources.

Digital Vehicle Forensics (DVF) is still an emerging research area. Compared with computer forensics, mobile forensics, cloud forensics, and industrial control system forensics, DVF is challenged by limited tool availability, proprietary vehicle architectures, safety-critical behavior, and the fact that vehicle data is often collected for functional or diagnostic purposes rather than for forensic purposes \cite{Altschaffel2020,Strandberg2022,AutomotiveBlackBox2023}. Existing research has demonstrated individual processes, component-level acquisitions, infotainment analyses, event-data-recorder (EDR) studies, and vehicle assistant application investigations. However, the literature still lacks a consolidated explanation of how the vehicle context differs from generic digital forensics and how those differences can guide evidence-source prioritization.

\textbf{Research gap.} Prior work has largely addressed one of four topics: investigation processes, in-vehicle component analysis, ecosystem artifacts, or forensic-readiness tools. These contributions are valuable but fragmented. What remains insufficiently articulated is a characteristic-level synthesis that (i) explains the vehicular forensic context, (ii) identifies recurring constraints across vehicles, components, and ecosystem sources, (iii) makes the derivation of such characteristics reproducible, and (iv) considers deliberate attempts to avoid, manipulate, or degrade forensic evidence. This gap matters because investigators need a defensible way to decide which sources to preserve first, which evidence to correlate, and which assumptions should be documented.

We addresses three research questions:
\begin{enumerate}
    \item Which capabilities can a vehicle, its components, and the vehicle ecosystem contribute to digital forensic investigations? \label{rq:capabilities} (RQ\ref{rq:capabilities})
    \item What characteristics can be assigned to vehicles in the domain of digital vehicle forensics, and how were those characteristics derived? \label{rq:characteristics} (RQ\ref{rq:characteristics})
    \item How can the characteristics support transparent, adversary-aware triage of vehicle-related digital evidence? \label{rq:triage} (RQ\ref{rq:triage})
\end{enumerate}

The contributions are:
\begin{enumerate}
    \item a definition of DVF and a system-level problem formulation that makes explicit the relevant variables, constraints, and assumptions;
    \item a structured review and synthesis protocol, including search dimensions, inclusion/exclusion criteria, and a coding process for deriving characteristics;
    \item a comparative positioning of the work against the closest prior studies and a set of eight characteristics with practical implications;
    \item a characteristic-driven triage procedure, analytical ablation table, limitations, failure cases, and adversarial/anti-forensic discussion.
\end{enumerate}

\section{Digital Vehicle Forensics and Problem Formulation}\label{sec:problem}

\subsection{What constitutes digital forensics in a vehicular context}
Digital forensics generally concerns the identification, preservation, acquisition, verification, analysis, interpretation, and reporting of digital evidence while maintaining evidential value, integrity, and chain of custody \cite{NIST80086,ISO27037,ISO27043}. In the vehicular context, the evidence space extends beyond a single seized device. It includes the vehicle as a whole, in-vehicle electronic control units (ECUs), infotainment and telematics systems, EDRs, diagnostic interfaces, paired smartphones, vehicle assistant applications, cloud back ends, charging infrastructure, insurance and fleet platforms, and physical traces produced by the vehicle.

DVF differs from generic computer forensics in three practical respects. First, many artifacts are functional rather than forensic: they were recorded for safety, diagnostics, control, warranty, or user convenience. Second, acquisition may influence safety-relevant state, especially during live or online examination. Third, interpretation often requires translating proprietary encodings into physical meaning, for example, mapping CAN frames to speed, braking, steering, or door-state signals.

\subsection{System model}
Let a vehicle-related incident be denoted by $e \in \event$, such as a crash, vehicle theft, suspected manipulation, cyberattack, hit-and-run, or disputed driver action. Let the vehicle and its ecosystem be modeled as
\begin{equation}
    V = (C,N,U,E,P,T),
\end{equation}
where $C=\{c_1,\ldots,c_n\}$ is the set of relevant components, $N$ is the set of in-vehicle and external communication links, $U$ is the set of drivers, passengers, maintainers, remote service operators, and other users, $E$ is the external ecosystem of phones, cloud services, infrastructure, and third-party systems, $P$ is the set of physical traces and constraints, and $T$ is the time base or set of time bases used by sources.

The candidate evidence-source set is
\begin{equation}
    \source = \{s_i \mid s_i \in C \cup E \cup P\}.
\end{equation}
Each source $s_i$ may yield an artifact $d_i$ if it is lawfully and technically acquired through an acquisition action $a_i$. Each source is associated with attributes
\begin{equation}
    \alpha_i = (v_i, b_i, r_i, \tau_i, \ell_i, q_i),
\end{equation}
where $v_i$ is volatility, $b_i$ is accessibility, $r_i$ is safety risk of acquisition, $\tau_i$ is expected integrity or trustworthiness, $\ell_i$ is legal/privacy cost, and $q_i$ is expected relevance to the investigative question. These attributes are not universal constants. They are case-specific assessments that must be documented.

The investigator seeks a reconstructed timeline or explanation $\hat{\timeline}$ of the incident:
\begin{equation}
    \hat{\timeline} = f(D,P,K), \quad D=\{d_i \mid s_i \in \source_A\},
\end{equation}
where $\source_A\subseteq\source$ is the set of acquired sources and $K$ is prior knowledge such as vehicle architecture, service documentation, physical laws, standards, and chain-of-custody records. A practical triage objective can be expressed as
\begin{equation}
\max_{\source_A \subseteq \source} \sum_{s_i \in \source_A} w_i \cdot I(d_i;e)
\end{equation}
subject to
\begin{align}
    r_i &\leq r_{\max},\label{6}\\
    \ell_i &\leq \ell_{\max},\label{7}\\
    \Delta state(s_i,a_i) &\leq \epsilon_i,\label{8}\\
    t(a_i) &\leq t_{\max}(v_i)\label{9},
\end{align}
with Equation \ref{6} being the \emph{safety constraint}, Equation \ref{7} the \emph{authorization and privacy constraint}, Equation \ref{8} the \emph{forensic-soundness constraint}, and Equation \ref{9} \text{volatility constraint}.

Here, $I(d_i;e)$ is the expected information about the incident, $w_i$ is a documented case-specific priority weight, and $\Delta state$ denotes the extent to which acquisition changes the source. This formulation is not intended to compute a universal optimum. It makes explicit the trade-offs that investigators already face: evidence value must be balanced against volatility, accessibility, safety, integrity, legality, privacy, and cost.

\subsection{Core vehicular forensic challenges}\label{sec:core-challenges}
\textbf{Isolation of connectivity.} A vehicle may communicate over cellular, Wi-Fi, Bluetooth, keyless-entry, GNSS, V2X, diagnostic, and charging interfaces. Isolation can preserve evidence by preventing remote wiping, synchronization, or command execution. However, isolation can also alter the vehicle state, interrupt ongoing logging, prevent back-end acquisition, or create safety risks. DVF therefore requires explicit documentation of connectivity state, isolation actions, timing, and consequences.

\textbf{Data extraction.} Extraction methods include diagnostic acquisition over OBD-II, Unified Diagnostic Services (UDS), Diagnostics over Internet Protocol (DoIP), USB or storage acquisition, mobile-device extraction, API-based cloud acquisition, JTAG/debug-interface acquisition, chip-off techniques, and manufacturer-assisted production. Each method differs in intrusiveness, repeatability, legal authority, and evidential risk. For example, infotainment chip-off analysis can provide rich artifacts but may be time-consuming and irreversible \cite{Jacobs2017,LeKhac2020}.

\textbf{Interpretation of vehicle-generated data.} Vehicle data is often not self-explanatory. CAN messages, diagnostic trouble codes, EDR parameters, timestamps, GPS points, mobile app caches, and cloud telemetry require context. Interpretation can be undermined by proprietary data formats, undocumented signal mappings, clock drift, different time bases, sensor uncertainty, unit conversions, and missing metadata. Reverse-engineering approaches such as CAN-D demonstrate the complexity of decoding CAN data when manufacturer signal definitions are unavailable \cite{Verma2021}.

\subsection{Boundary conditions and assumptions}\label{sec:assumptions}
This article focuses on road vehicles and vehicle-related digital evidence. It does not assume that all vehicles are connected, that all OEMs expose the same artifacts, or that evidence is always obtainable without manufacturer cooperation. It assumes lawful authority, a documented chain of custody, and compliance with safety requirements. It further assumes that investigators should prefer the least intrusive acquisition method that can answer the question, unless volatility or imminent loss of evidence justifies a more intrusive method.

\section{Review Method and Characteristic Synthesis}\label{sec:method}

\subsection{Review design}
The work uses a structured scoping review rather than a full systematic review. The goal is to identify recurring capabilities and constraints that define DVF, not to estimate the prevalence of a phenomenon or compute pooled performance statistics. The reporting structure follows scoping-review principles by making the search dimensions, selection criteria, and synthesis process explicit \cite{PRISMAScR2018}.

\begin{table*}[!t]
\caption{Review and synthesis protocol used to identify capabilities and derive DVF characteristics.}
\label{tab:review-protocol}
\centering
\small
\begin{tabularx}{\textwidth}{lY}
\toprule
\textbf{Element} & \textbf{Protocol used in this article} \\
\midrule
Academic search spaces & IEEE Xplore, ACM Digital Library, ScienceDirect, SpringerLink, Scopus/Google Scholar, and backward/forward snowballing from key DVF papers. \\
Query families & (``vehicle'' OR ``automotive'' OR ``connected car'' OR ``smart vehicle'') AND (``digital forensics'' OR ``forensic'' OR ``digital evidence'' OR ``event reconstruction''); (``CAN'' OR ``OBD'' OR ``UDS'' OR ``DoIP'' OR ``EDR'' OR ``infotainment'' OR ``telematics'') AND (``forensic'' OR ``evidence''); (``vehicle assistant app'' OR ``vehicle cloud'' OR ``automotive back end'') AND (``forensic'' OR ``investigation''). \\
Non-academic sources considered & Standards, regulation documents, practitioner material, training/tool descriptions, manufacturer-facing cybersecurity rules, and legal/practitioner discussions were considered when they clarified acquisition, admissibility, or artifact availability. They were not used as sole evidence for a characteristic unless supported by technical literature or case material. \\
Inclusion criteria & A source was included when it described a vehicle, component, or ecosystem artifact; introduced or evaluated a forensic process, method, tool, data source, architecture, or standard; or provided a technically relevant limitation for evidence acquisition, preservation, interpretation, or reporting. \\
Exclusion criteria & Sources were excluded when they discussed general automotive security without forensic relevance, contained only marketing claims, lacked sufficient technical detail, duplicated a more complete version, or addressed non-road-vehicle domains without a transferable forensic implication. \\
Coding fields & Evidence source, acquisition interface, volatility, data semantics, user attribution, physical interpretation, component dependency, ecosystem dependency, safety impact, privacy/legal issue, anti-forensic risk, tool or standard dependency, and stated limitation. \\
Characteristic retention & A candidate characteristic was retained when it appeared across multiple independent sources or when a case study demonstrated a distinct forensic action or constraint that could not be adequately explained by generic computer forensics alone. \\
\bottomrule
\end{tabularx}
\end{table*}

\subsection{Derivation of the eight characteristics}\label{sec:derivation}
The characteristics were not copied verbatim from a single publication. They were synthesized by coding the included sources for recurring forensic capabilities and constraints, grouping similar codes, and retaining candidate groups that changed how an investigator would preserve, acquire, interpret, or validate evidence. Table \ref{tab:derivation} summarizes the resulting characteristics and explains the derivation logic.

\begin{table*}[!t]
\caption{Derived characteristics and their forensic interpretation.}
\label{tab:derivation}
\centering
\scriptsize
\begin{tabularx}{\textwidth}{c l Y Y}
\toprule
\textbf{ID} & \textbf{Characteristic} & \textbf{Definition in DVF} & \textbf{Why it was retained} \\
\midrule
$C_1$ & Multiple users & A vehicle may be used or influenced by several drivers, passengers, owners, app users, service personnel, or remote account holders. & Recurrent user-attribution problem in infotainment, smartphone, car-sharing, assistant-app, and cloud artifacts. \\
$C_2$ & Massively networked & Vehicle data is distributed across internal buses, diagnostic interfaces, telematics, mobile devices, cloud services, infrastructure, and third-party platforms. & Recurrent need to correlate vehicle, phone, cloud, and network data; single-source analysis is often incomplete. \\
$C_3$ & Cyber-physical system & Digital traces correspond to physical states and actions such as motion, braking, steering, acceleration, door state, location, or impact. & Enables physical plausibility validation and distinguishes DVF from many purely digital investigations. \\
$C_4$ & Dependencies between components & Components may require other ECUs, cryptographic material, gateways, sensors, clocks, or back-end services to operate or be interpreted. & Affects live acquisition, bench testing, and attribution of where the same event may be recorded. \\
$C_5$ & Functional data & Many artifacts exist for diagnostics, control, safety, warranty, or user services rather than for forensic logging. & Explains semantic gaps, missing forensic metadata, proprietary formats, and limited evidential context. \\
$C_6$ & Safety implications & Acquisition, analysis, or manipulation may affect safety-critical systems or involve a safety-relevant incident. & Live analysis and reinstallation can create hazards; safety constrains forensic actions. \\
$C_7$ & Accessibility & Vehicles and components may be physically remote, integrated, locked, encrypted, cloud-dependent, or proprietary. & Determines whether evidence can be acquired before volatility, overwriting, or legal delay reduces value. \\
$C_8$ & Limited abstraction & Compared with general IT, many vehicle artifacts lack standardized abstraction layers, logs, names, file systems, and public schemas. & Necessitates reverse engineering, OEM cooperation, and careful uncertainty reporting. \\
\bottomrule
\end{tabularx}
\end{table*}

\section{Related Work and Positioning}\label{sec:related}
This section reorganizes the related work into four areas and explains how the present synthesis differs from them.

\subsection{Processes for digital vehicle forensics}
Kuhlmann et al. studied how IT incidents can affect automotive safety and driver reactions \cite{Kuhlmann2015}. Mansor discussed security, privacy, and process challenges in automotive systems, including data availability and privacy \cite{Mansor2017}. Altschaffel et al. proposed a five-step process that distinguishes strategic preparation, operational preparation, data collection, examination, and analysis \cite{Altschaffel2017}. Gomez Buquerin et al. proposed a generalized automotive-forensics process with forensic readiness, data collection, data analysis, and documentation, and demonstrated the process through diagnostic communication with a modern vehicle \cite{Gomez2021c}. These process papers clarify investigation phases but do not provide a characteristic-level taxonomy that explains why certain vehicle evidence sources behave differently from generic digital sources.

\subsection{In-depth vehicle and component investigations}
Component-level studies show that DVF can recover valuable evidence but also reveal acquisition and interpretation barriers. Kiltz et al. investigated automotive data types and analyses, including debugging interfaces \cite{Klitz2009}. Koscher et al. experimentally analyzed a modern automobile and showed the security implications of internal networks and controller interactions \cite{Koscher2010}. Hoppe et al. reconstructed a driven route using automotive communication data \cite{Hoppe2012}. Jacobs et al. extracted artifacts from a Volkswagen infotainment system using embedded-forensics techniques \cite{Jacobs2017}. Vandiver and Anderson analyzed Berla iVe acquisitions from Ford SYNC systems \cite{Vandiver2018}. Le-Khac et al. showed that classical forensic tools do not reliably analyze all tested automotive file systems, motivating DVF-specific tooling \cite{LeKhac2020}. Gomez Buquerin and Hof analyzed a Tesla Autopilot file system \cite{Gomez2022}, and Kurachi et al. evaluated manipulation concerns for EDR data \cite{Kurachi2022}.

\subsection{Vehicle ecosystem, mobile, cloud, and third-party artifacts}
Vehicles increasingly create evidence outside the car. Attenberger identified automotive forensic data sources beyond isolated in-vehicle storage \cite{Attenberger2020}. Ebbers et al. analyzed vehicle assistant applications and showed that smartphone and manufacturer back-end data can reconstruct driver activities \cite{Ebbers2021}. Sumaila and Bahsi investigated automotive maintenance applications and found differences in the quantity and quality of artifacts \cite{Sumaila2022}. Recent work has extended the ecosystem view to API-based vehicle cloud acquisition and usage-based insurance data \cite{Ebbers2024,Onik2024}. These studies support the ``massively networked'' and ``accessibility'' characteristics and show why industry reports, practitioner documentation, and legal access mechanisms must be considered in addition to academic papers.

\subsection{Forensic readiness, standards, and comparative surveys}
Strandberg et al. systematically reviewed automotive digital forensics and classified challenges, data sources, communication, software, hardware, algorithms, cryptography, processes, infrastructure, and virtualization \cite{Strandberg2022}. Altschaffel compared automotive, desktop-IT, and industrial-control-system forensic influence factors \cite{Altschaffel2020}. The Automotive BlackBox work proposed forensic requirements and a standardization-oriented architecture for automotive forensic-enabled vehicles \cite{AutomotiveBlackBox2023}. Li et al. proposed public-auditing mechanisms for in-vehicle forensic data in connected and automated vehicles \cite{Li2024}. These works provide broad coverage or forensic-by-design proposals. The present article differs by deriving a compact set of characteristics and translating them into triage, assumptions, adversarial risks, and failure cases.

\begin{table*}[!t]
\caption{Positioning against closest prior work.}
\label{tab:comparison}
\centering
\scriptsize
\begin{tabularx}{\textwidth}{l c c c c c Y}
\toprule
\textbf{Work} & \textbf{Process} & \textbf{Component} & \textbf{Ecosystem} & \textbf{Taxonomy} & \textbf{Adversarial} & \textbf{Main distinction from this article} \\
\midrule
Altschaffel et al. \cite{Altschaffel2017} & \cmark &  &  &  &  & Proposes a DVF process; does not derive recurring characteristics or adversarial triage implications. \\
Gomez Buquerin et al. \cite{Gomez2021c} & \cmark & \cmark &  &  &  & Demonstrates a generalized process and diagnostic acquisition; does not provide a characteristic taxonomy. \\
Le-Khac et al. \cite{LeKhac2020} &  & \cmark &  &  &  & Provides a smart-vehicle case study and tool challenge analysis; narrower component focus. \\
Ebbers et al. \cite{Ebbers2021,Ebbers2024} &  &  & \cmark &  & partially & Focuses on apps, back ends, and API acquisition; supports but does not generalize to all DVF characteristics. \\
Strandberg et al. \cite{Strandberg2022} & \cmark & \cmark & \cmark & \cmark & partially & Broad systematic review; categories are research areas and challenges rather than investigation characteristics. \\
Automotive BlackBox \cite{AutomotiveBlackBox2023} &  & \cmark & \cmark &  & \cmark & Forensic-by-design architecture and requirements; this article focuses on current-investigation characteristics and triage. \\
This article & \cmark & \cmark & \cmark & \cmark & \cmark & Defines DVF, formalizes the triage problem, derives eight characteristics, and adds assumptions, limitations, failure cases, and anti-forensic considerations. \\
\bottomrule
\end{tabularx}
\end{table*}

\section{Characteristics of Digital Vehicle Forensics}\label{sec:characteristics}
Table \ref{tab:characteristics-related-work} maps representative sources to the eight characteristics. A check mark indicates that the source provides evidence, an example, or a challenge supporting the characteristic. The mapping is not a claim that each source explicitly named the characteristic; it documents how the characteristic was synthesized from the coded literature.

\begin{table*}[!t]
\caption{Representative literature support for the eight characteristics.}
\label{tab:characteristics-related-work}
\centering
\scriptsize
\begin{tabularx}{\textwidth}{Y c c c c c c c c}
\toprule
\textbf{Source} & \textbf{$C_1$} & \textbf{$C_2$} & \textbf{$C_3$} & \textbf{$C_4$} & \textbf{$C_5$} & \textbf{$C_6$} & \textbf{$C_7$} & \textbf{$C_8$} \\
\midrule
Kuhlmann et al. \cite{Kuhlmann2015} & & & & & & \cmark & & \\
Mansor \cite{Mansor2017} & & & & & & & \cmark & \\
Altschaffel et al. \cite{Altschaffel2017} & & \cmark & & \cmark & & & & \cmark \\
Gomez Buquerin et al. \cite{Gomez2021c} & & \cmark & & \cmark & \cmark & & \cmark & \cmark \\
Kiltz et al. \cite{Klitz2009} & & & \cmark & \cmark & & & & \cmark \\
Koscher et al. \cite{Koscher2010} & & \cmark & \cmark & \cmark & \cmark & \cmark & \cmark & \\
Hoppe et al. \cite{Hoppe2012} & & \cmark & \cmark & & \cmark & \cmark & \cmark & \\
Jacobs et al. \cite{Jacobs2017} & \cmark & \cmark & & & \cmark & \cmark & \cmark & \cmark \\
Vandiver and Anderson \cite{Vandiver2018} & \cmark & \cmark & \cmark & & \cmark & \cmark & \cmark & \\
Le-Khac et al. \cite{LeKhac2020} & \cmark & \cmark & & & \cmark & \cmark & \cmark & \cmark \\
Gomez Buquerin and Hof \cite{Gomez2022} & & \cmark & \cmark & \cmark & \cmark & & \cmark & \cmark \\
Sumaila and Bahsi \cite{Sumaila2022} & \cmark & \cmark & \cmark & \cmark & \cmark & & & \\
Attenberger \cite{Attenberger2020} & & \cmark & & & & & \cmark & \\
Ebbers et al. \cite{Ebbers2021,Ebbers2024} & \cmark & \cmark & & & \cmark & & \cmark & \cmark \\
Kurachi et al. \cite{Kurachi2022} & & \cmark & \cmark & \cmark & \cmark & \cmark & & \\
Strandberg et al. \cite{Strandberg2022} & \cmark & \cmark & \cmark & \cmark & \cmark & \cmark & \cmark & \cmark \\
\bottomrule
\end{tabularx}
\end{table*}

\subsection{Multiple users -- $C_1$}
Vehicles can be shared by households, employees, rental users, car-sharing subscribers, passengers, maintenance personnel, remote app users, and service providers. User attribution is therefore a central DVF challenge. Infotainment systems may store paired Bluetooth devices, call logs, navigation destinations, media identifiers, contacts, or smartphone-integration artifacts. Vehicle assistant apps and cloud services may link actions to accounts or devices, while seat, climate, and profile settings may suggest but not prove individual use.

For investigations, $C_1$ means that a vehicle artifact should not automatically be attributed to the registered owner or the last known driver. A Bluetooth identifier may show that a phone was paired; it does not by itself prove who drove the vehicle at a particular time. Conversely, separating multiple users can help exclude irrelevant artifacts or identify co-offenders. The implication is that DVF reports should distinguish device attribution, account attribution, physical presence, and driver attribution.

\subsection{Massively networked -- $C_2$}
Vehicles are networked internally and externally. Internal buses and gateways connect ECUs, infotainment, telematics, comfort systems, safety systems, and diagnostic interfaces. External links may include cellular connectivity, Wi-Fi, Bluetooth, GNSS, V2X, manufacturer back ends, charging infrastructure, insurance platforms, and third-party services. The global sale of vehicles with embedded telematics increased substantially during the 2010s \cite{Statista2020}, and vehicle-centric connected services have been projected as economically significant \cite{Statista2019}.

This characteristic enables correlation across sources but creates a data-volume and data-location challenge. Gomez Buquerin et al. found more than 100 logical addresses reachable through a modern vehicle diagnostic interface \cite{Gomez2021c}. We also performed an illustrative network scan of a Tesla Model X by connecting to the internal WLAN and scanning IPv4 endpoints and ports. The scan identified 96 IPv4 endpoints and showed dominant TCP and TLS traffic; the protocol hierarchy is given in Appendix \ref{app:nmap}. This example is not a statistically representative dataset. Its purpose is to illustrate the internal networked nature of a modern vehicle and to motivate why triage and correlation are needed.

Networked evidence can also exist outside the vehicle. Shodan queries have been used to identify exposed vehicle-connected services, such as electric-vehicle chargers and GPS-tracker interfaces \cite{Jarvis2022}. From a forensic viewpoint, $C_2$ requires investigators to decide whether the vehicle, paired devices, cloud services, infrastructure, or third-party providers should be preserved first.

\subsection{Cyber-physical system -- $C_3$}
Vehicles are cyber-physical systems because digital states correspond to physical behavior. Speed, braking, steering angle, acceleration, seat-belt state, impact events, door state, and geolocation are not merely software variables; they describe physical actions and constraints. This property enables plausibility checks that are not available in many purely digital investigations. For instance, an EDR report showing a physically impossible acceleration profile should be treated as suspect or at least requiring explanation.

The cyber-physical character also creates interpretation risk. A sensor value may be valid but misinterpreted if units, sampling intervals, coordinate systems, or event triggers are misunderstood. Nilsson and Larson emphasized the value of combining physical and digital evidence in vehicle environments \cite{Nilsson2008a}. EDR studies show that vehicle speed, brake state, accelerator pedal state, and related parameters can support reconstruction, but the trustworthiness of such data must still be assessed \cite{Daily2008,Lee2022,Kurachi2022}.

\subsection{Dependencies between components -- $C_4$}
Vehicle components may depend on other components, gateways, clocks, cryptographic material, sensors, or manufacturer services. A removed ECU may not boot or may not expose data without the correct harness, gateway, immobilizer state, time source, or security access. Hardware-in-the-loop systems can simulate some dependencies for testing and acquisition, but they are not always available to forensic laboratories \cite{Joshi2017}.

For investigations, $C_4$ affects both acquisition and interpretation. A diagnostic trouble code in one ECU may be caused by a fault or manipulation elsewhere. A telematics unit may hold data whose meaning depends on cloud synchronization. A component dependency may also provide corroboration: if several controllers observed the same event, their timestamps and state changes can be compared.

\subsection{Functional data -- $C_5$}
Most vehicle data is collected for functional reasons: diagnostics, safety, control, warranty, maintenance, user convenience, or service delivery. Standards such as ISO 26262 address functional safety \cite{ISO26262}, while UNECE Regulation No. 155 and ISO/SAE 21434 address cybersecurity management and engineering \cite{UNECE2020,ISO21434}. These standards and regulations do not by themselves provide a complete forensic logging standard for all vehicle artifacts. Recent forensic-by-design proposals therefore argue for additional requirements, data formats, integrity properties, and standardized extraction paths \cite{AutomotiveBlackBox2023,Li2024}.

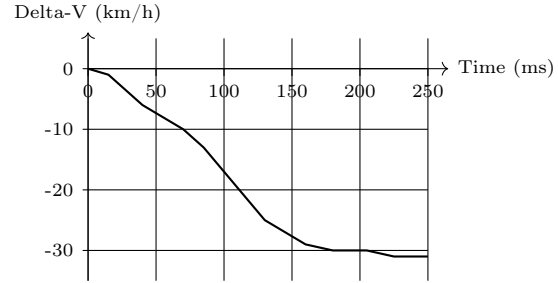
\begin{figure}[!t]
\centering
\begin{tikzpicture}[x=0.018cm,y=0.08cm,font=\scriptsize]
    \draw[->] (0,-35) -- (0,6) node[above] {Delta-V (km/h)};
    \draw[->] (0,0) -- (265,0) node[right] {Time (ms)};
    \foreach \x in {0,50,100,150,200,250} {
        \draw (\x,0) -- (\x,-1.2) node[below] {\x};
        \draw[thin] (\x,-35) -- (\x,5);
    }
    \foreach \y in {-30,-20,-10,0} {
        \draw (0,\y) -- (-4,\y) node[left] {\y};
        \draw[thin] (0,\y) -- (250,\y);
    }
    \draw[thick] plot coordinates {(0,0) (15,-1) (25,-3) (40,-6) (55,-8) (70,-10) (85,-13) (100,-17) (115,-21) (130,-25) (145,-27) (160,-29) (180,-30) (205,-30) (225,-31) (250,-31)};
\end{tikzpicture}
\caption{Recreated longitudinal delta-V plot from a Tesla Model 3 EDR-style report. The figure illustrates functional data that can support physical plausibility checks but requires contextual interpretation.}
\label{fig:tesla-edr}
\end{figure}

Figure \ref{fig:tesla-edr} shows a recreated, readable version of an EDR-style longitudinal delta-V plot. Such data can support crash reconstruction, but it does not necessarily explain intent, cyberattack causality, or user identity. Functional data must therefore be translated into forensic propositions with appropriate uncertainty.

\subsection{Safety implications -- $C_6$}
DVF may concern safety-relevant incidents, and forensic actions may themselves affect safety. Live acquisition from a running vehicle can change timing, bus load, controller state, or driver information. Kuhlmann et al. showed that IT-related incidents can influence automotive safety, highlighting the interaction between cybersecurity and safety \cite{Kuhlmann2015}. Consequently, investigators should avoid live actions on an operating vehicle unless the safety case is understood and documented.

Post-mortem acquisition also has safety implications. If a component is removed, powered externally, altered, or reinstalled, its correct operation must be considered. This is especially important for safety-critical components and for vehicles that may later be returned to service.

\subsection{Accessibility -- $C_7$}
Vehicles are usually not in forensic laboratories. They may be at crash scenes, private homes, dealerships, impound lots, repair shops, or remote locations. Components may be physically difficult to access, encrypted, locked behind secure gateways, or dependent on OEM cooperation. Volatile memory, cloud-retention windows, and synchronization behavior can make delay harmful. Chip-off and embedded acquisition can recover data but may be destructive or irreversible \cite{Jacobs2017,LeKhac2020}. Practitioner sources show that vehicle forensic work often requires disassembly, specialist tools, and training \cite{Digitpol2023,ABForensics2021}.

The implication is that DVF planning must document what was accessible, when it was accessible, why more intrusive methods were or were not used, and how volatility was mitigated.

\subsection{Limited abstraction -- $C_8$}
Unlike general-purpose computers, vehicles often lack standardized forensic abstractions such as uniform logs, stable file paths, public schemas, and consistent time sources. Frameworks such as AUTOSAR and automotive-grade Linux provide abstraction for development, but they do not guarantee forensic access or uniform evidence semantics across manufacturers and components. Proprietary CAN signal definitions are a clear example: without manufacturer mappings, raw traffic may be hard to interpret \cite{Verma2021}.

$C_8$ requires investigators to report uncertainty. A decoded signal, inferred user action, or reconstructed route should be linked to the method used, its validation status, and its limitations. Where reverse engineering is required, reproducibility and peer review become especially important.

\begin{table*}[!t]
\caption{Characteristics and practical use in investigations.}
\label{tab:characteristics}
\centering
\small
\begin{tabularx}{\textwidth}{c l Y}
\toprule
\textbf{ID} & \textbf{Characteristic} & \textbf{Practical application} \\
\midrule
$C_1$ & Multiple users & Separate device, account, passenger, and driver attribution; avoid assuming that vehicle ownership equals driver identity. \\
$C_2$ & Massively networked & Preserve and correlate vehicle, phone, cloud, infrastructure, and network traces; handle big-data and jurisdictional issues. \\
$C_3$ & Cyber-physical system & Validate digital evidence against physical constraints such as speed, acceleration, braking, location, and impact dynamics. \\
$C_4$ & Dependencies between components & Account for component dependencies during bench testing, live acquisition, and cross-controller corroboration. \\
$C_5$ & Functional data & Translate diagnostic, safety, warranty, or service data into forensic propositions while documenting semantic uncertainty. \\
$C_6$ & Safety implications & Avoid acquisition actions that can create hazards; document safety cases for live analysis and component reinstallation. \\
$C_7$ & Accessibility & Prioritize volatile or remote sources; document why inaccessible, encrypted, cloud-dependent, or intrusive sources were not acquired. \\
$C_8$ & Limited abstraction & Use reverse engineering, OEM cooperation, validation, and uncertainty reporting when standard abstractions are absent. \\
\bottomrule
\end{tabularx}
\end{table*}

\section{Characteristic-Driven Triage Procedure}\label{sec:triage}
Reviewer feedback requested clearer algorithmic structure, assumptions, complexity, ablation, and parameter justification. The contribution of this article is not a classifier or signal-processing algorithm; nevertheless, the characteristics can be operationalized as a reproducible triage procedure. Algorithm \ref{alg:triage} converts the characteristics into a documented evidence-source prioritization workflow.

\begin{algorithm}[!t]
\caption{Characteristic-driven DVF triage}
\label{alg:triage}
\begin{algorithmic}[1]
\REQUIRE Incident question $e$, candidate sources $\source$, constraints $(r_{\max},\ell_{\max},\epsilon,t_{\max})$, characteristic set $C_1\ldots C_8$
\ENSURE Prioritized acquisition plan $P_A$ and documented assumptions
\STATE Preserve the scene; record time, power state, connectivity state, odometer, visible damage, and legal authority.
\STATE Identify candidate vehicle, component, mobile, cloud, infrastructure, and physical sources $\source$.
\FORALL{$s_i \in \source$}
    \STATE Assign documented attributes $\alpha_i=(v_i,b_i,r_i,\tau_i,\ell_i,q_i)$.
    \STATE Mark applicable characteristics $C_j(s_i)$, $j \in \{1,\ldots,8\}$.
    \STATE Exclude or delay $s_i$ if safety risk, legal authority, or state-change constraints are not satisfied.
    \STATE Estimate priority from relevance, volatility, accessibility, integrity, and corroboration value.
\ENDFOR
\STATE Acquire sources in priority order using the least intrusive method that can answer the question.
\STATE Verify integrity by hashing, chain-of-custody records, tool validation, and repeatability when possible.
\STATE Correlate artifacts across users, time bases, components, ecosystem sources, and physical constraints.
\STATE Identify contradictions, missing sources, anti-forensic indicators, and assumptions.
\STATE Report $\hat{\timeline}$ with evidence, uncertainty, limitations, and alternative explanations.
\end{algorithmic}
\end{algorithm}

\subsection{Complexity and scalability}
Let $n=|\source|$ be the number of candidate sources, $m=8$ the number of characteristics, and $k$ the number of artifact records acquired from a source after parsing. Characteristic annotation is $O(nm)$, which is linear in the number of sources because $m$ is fixed. Pairwise correlation across sources is $O(n^2 k)$ in the naive case, but practical investigations reduce this by filtering on time windows, event type, location, and relevance. Memory cost is $O(nk)$ for artifact metadata, excluding raw forensic images. The framework therefore scales conceptually with the number of sources and artifacts, not with model hyperparameters.

\subsection{Metric selection and validation rationale}
DVF triage should not be evaluated only by speed. Useful evaluation dimensions are: (i) coverage of plausible evidence sources, (ii) preservation of volatile data, (iii) integrity and repeatability of acquisition, (iv) corroboration across independent sources, (v) safety risk, (vi) privacy/legal proportionality, and (vii) explanatory value for the incident question. These metrics align with forensic goals rather than machine-learning accuracy. The Tesla network scan and EDR plot in this article are illustrative case material; they are not a dataset for statistical performance claims. Consequently, mean, variance, confidence intervals, runtime benchmarks, noise-injection results, and hyperparameter sensitivity plots are not reported. Future empirical studies should evaluate these dimensions across multiple manufacturers, model years, and incident types.

\subsection{Analytical ablation}
Because the contribution is a characteristic framework, the appropriate ablation is analytical: what investigative capability is lost if a characteristic is ignored? Table \ref{tab:ablation} summarizes the effect of omitting each characteristic.

\begin{table*}[!t]
\caption{Analytical ablation: effect of ignoring each characteristic.}
\label{tab:ablation}
\centering
\small
\begin{tabularx}{\textwidth}{c Y Y}
\toprule
\textbf{Omitted} & \textbf{Likely loss} & \textbf{Example consequence} \\
\midrule
$C_1$ & Weak user attribution & A paired phone is treated as driver proof without considering passengers or account sharing. \\
$C_2$ & Incomplete source coverage & Cloud, app, infrastructure, or diagnostic data is missed even though it could corroborate the vehicle. \\
$C_3$ & No physical plausibility check & Implausible speed, acceleration, or location data is accepted without validation. \\
$C_4$ & Misinterpreted component evidence & A fault in one ECU is analyzed without considering gateway, clock, or sensor dependencies. \\
$C_5$ & Overstated semantics & Diagnostic or functional data is interpreted as direct evidence of intent or cyberattack causality. \\
$C_6$ & Unsafe acquisition & Live examination interferes with safety-relevant systems or reinstallation occurs without validation. \\
$C_7$ & Lost volatile or remote evidence & Cloud retention windows, volatile memory, or inaccessible components are not prioritized. \\
$C_8$ & Undocumented uncertainty & Reverse-engineered signals or proprietary formats are reported as if they were standardized logs. \\
\bottomrule
\end{tabularx}
\end{table*}

\subsection{Parameter selection and reproducibility}
Algorithm \ref{alg:triage} contains case-specific priorities rather than universal hyperparameters. If an organization assigns numerical weights to relevance, volatility, safety, or accessibility, those weights should be justified in the case file and sensitivity should be checked by re-ranking sources under plausible alternative weights. For reproducibility, investigators should retain: search warrants or legal authority, tool versions, acquisition commands, vehicle state, connectivity state, time synchronization notes, hash values, parsed-artifact schemas, reverse-engineering notes, and reasons for excluding sources.

\section{Adversarial Perspective and Anti-Forensics}\label{sec:adversarial}
DVF cannot assume that evidence is passively produced and preserved. A driver, owner, attacker, insider, or remote service operator may try to prevent collection, manipulate artifacts, or create misleading traces. Automotive security studies demonstrate that remote and local compromise of vehicle systems is realistic \cite{Koscher2010,Checkoway2011,MillerValasek2015}. EDR-focused research also shows that crash-related data can be a target for manipulation \cite{Kurachi2022}. The Automotive BlackBox paper explicitly motivates forensic-by-design partly because vehicle attacks may erase or tamper with evidence \cite{AutomotiveBlackBox2023}.

\begin{table*}[!t]
\caption{Adversarial and anti-forensic considerations in DVF.}
\label{tab:antiforensics}
\centering
\scriptsize
\begin{tabularx}{\textwidth}{Y Y c Y}
\toprule
\textbf{Anti-forensic tactic} & \textbf{Vehicle manifestation} & \textbf{Affected characteristics} & \textbf{Forensic response} \\
\midrule
Evidence deletion or reset & Infotainment factory reset, mobile-app cache deletion, removal of paired devices, cloud-token revocation. & $C_1,C_2,C_5,C_7$ & Acquire volatile and ecosystem sources early; request back-end data; compare phone, vehicle, and cloud traces. \\
Connectivity manipulation & Disconnecting telematics, SIM removal, RF shielding by suspect, GNSS jamming or spoofing, disabling Bluetooth/Wi-Fi. & $C_2,C_3,C_7$ & Document connectivity state; preserve logs from multiple time bases; compare GNSS with physical route and external sources. \\
Message injection or replay & CAN injection, diagnostic abuse, replayed UDS sessions, spoofed sensor states. & $C_2,C_3,C_4,C_6$ & Correlate multiple ECUs; inspect diagnostic session traces; validate physical plausibility and timing consistency. \\
Firmware or configuration tampering & ECU reflashing, unauthorized coding changes, modified telematics or infotainment firmware. & $C_4,C_5,C_6,C_8$ & Preserve firmware versions, signatures, calibration identifiers, update logs, and OEM records. \\
Clock and timestamp manipulation & Incorrect vehicle clock, app clock drift, inconsistent cloud/vehicle time bases. & $C_2,C_5,C_8$ & Establish time offsets; compare independent clocks; report uncertainty and time-window assumptions. \\
Data poisoning through user behavior & Phone swapping, account sharing, deliberate route creation, fake GPS apps, staged trips. & $C_1,C_2,C_3$ & Separate account, device, and driver attribution; correlate with physical evidence and independent ecosystem records. \\
Physical destruction or access denial & Damaged ECUs, removed memory, destroyed phone, vehicle inaccessible at private or remote location. & $C_6,C_7$ & Prioritize less fragile sources, obtain legal authority promptly, use manufacturer or cloud records where available. \\
\bottomrule
\end{tabularx}
\end{table*}

The adversarial view changes the evidential standard. It is not sufficient to show that a data item exists; the investigator must ask how it could have been created, deleted, altered, or misinterpreted. Countermeasures include early preservation, independent corroboration, physical plausibility checks, tool validation, hash-based integrity verification, documentation of acquisition side effects, and explicit reporting of uncertainty.

\section{Discussion, Limitations, and Failure Cases}\label{sec:discussion}

\subsection{Generalization of the characteristics}
No single characteristic is unique to vehicles. Multiple users exist in smart homes, networked evidence exists in cloud and IoT systems, and cyber-physical constraints exist in industrial control systems. The contribution is the combination of all eight characteristics in the vehicle context. A modern vehicle can be a shared consumer object, a safety-critical cyber-physical system, an embedded network, a cloud-connected service platform, and a proprietary diagnostic environment at the same time. This combination explains why direct transfer of computer-forensic processes is insufficient.

The characteristics are expected to apply broadly to modern vehicles, but not uniformly. A legacy vehicle may provide limited telematics or app data, reducing $C_2$ and ecosystem evidence. A highly centralized software-defined vehicle may reduce some component dependencies but increase cloud, account, and software-update dependencies. Therefore, the characteristics should be treated as prompts for investigation rather than as assumptions that every vehicle contains every artifact.

\subsection{Practical implications}
The characteristics help investigators prioritize work. In a robbery involving a vehicle, $C_1$ directs attention to user attribution; $C_2$ expands the source list to phone, cloud, tolling, charging, and infrastructure records; $C_3$ supports route plausibility; $C_5$ prevents overclaiming diagnostic data as intent; and $C_7$ highlights retention and access windows. In a suspected vehicle cyberattack, $C_4$, $C_6$, and $C_8$ emphasize component dependencies, safety constraints, and reverse-engineering uncertainty.

\subsection{Failure cases and boundary conditions}
DVF may underperform or fail when no relevant digital source exists, when sources have been overwritten, when cloud retention has expired, when legal authority does not permit access, when encryption prevents acquisition, when manufacturer cooperation is unavailable, or when physical damage destroys storage. Interpretation can fail when proprietary encodings are unknown, clocks are inconsistent, or multiple plausible drivers and devices fit the evidence. Anti-forensic actions such as factory resets, fake GPS applications, remote wiping, or firmware tampering can further reduce confidence.

\subsection{Limitations}
This article has four main limitations. First, the review is a structured scoping review and not a full systematic literature review with exhaustive database counts. Second, the Tesla Model X network scan and EDR-style plot are illustrative; they do not support statistical claims about all manufacturers, model years, or configurations. Third, practitioner literature, industry reports, and legal cases can be important in DVF but are not always public, technically detailed, or peer reviewed. Fourth, the framework does not implement an automated parser, classifier, or detector; therefore, empirical ablation, runtime benchmarks, noise-injection experiments, and hyperparameter plots are outside its claim boundary. These omissions are not hidden; they define the scope and motivate future empirical work.

\subsection{Future work}
Future research should test the characteristics across multi-OEM datasets, controlled acquisition settings, and realistic adversarial scenarios. Useful studies would include controlled clock-skew experiments, data-loss and noise perturbation, back-end retention comparisons, safety-case analysis for live acquisition, and validation of characteristic-driven triage against completed case studies. Standardized artifact schemas, forensic-by-design logging, and public benchmark datasets would improve reproducibility and allow statistical validation.

\section{Conclusion}\label{sec:conclusion}
This article revised and strengthened the contribution by defining DVF, formalizing the vehicle-forensic triage problem, clarifying the review protocol, explaining how the eight characteristics were derived, comparing the work with close prior studies, and adding adversarial, limitation, and failure-case analysis. Vehicles can contribute valuable evidence through in-vehicle components, physical behavior, user interactions, diagnostic systems, mobile applications, cloud services, and infrastructure. However, this evidence is constrained by user attribution, networking, cyber-physical semantics, component dependencies, functional data, safety, accessibility, and limited abstraction. Treating these as explicit characteristics helps investigators plan acquisitions, explain uncertainty, identify missing sources, and communicate findings in a more defensible way.

\newpage
\appendix
\section{Illustrative Nmap Scan}\label{app:nmap}
Table \ref{tab:protocol_hier} reports the protocol hierarchy observed in the illustrative Tesla Model X Nmap scan discussed in Section \ref{sec:characteristics}. The scan is included as a case example of $C_2$ and should not be interpreted as representative of all vehicles.

\begin{table*}[!t]
\caption{Protocol hierarchy of the illustrative Nmap scan with a Tesla Model X.}
\label{tab:protocol_hier}
\centering
\small
\begin{tabularx}{\textwidth}{Y r r}
\toprule
\textbf{Protocol} & \textbf{Percent packets} & \textbf{Percent bytes} \\
\midrule
Ethernet & 100.00 & 11.15 \\
Internet Protocol Version 4 & 99.91 & 15.91 \\
User Datagram Protocol & 0.78 & 0.05 \\
Network Time Protocol & 0.06 & 0.02 \\
NetBIOS Name Service & 0.17 & 0.10 \\
NetBIOS Datagram Service & 0.02 & 0.03 \\
SMB (Server Message Block Protocol) & 0.02 & 0.01 \\
SMB MailSlot Protocol & 0.02 & 0.00 \\
Microsoft Windows Browser Protocol & 0.02 & 0.00 \\
Multicast Domain Name System & 0.24 & 0.20 \\
Domain Name System & 0.17 & 0.09 \\
Data & 0.11 & 0.03 \\
Transmission Control Protocol & 99.05 & 72.36 \\
Transport Layer Security & 4.22 & 32.04 \\
NetBIOS Session Service & 0.09 & 0.06 \\
SMB (Server Message Block Protocol) & 0.09 & 0.06 \\
SMB Pipe Protocol & 0.04 & 0.00 \\
Microsoft Windows Lanman Remote API Protocol & 0.04 & 0.01 \\
Hypertext Transfer Protocol & 0.05 & 0.22 \\
Online Certificate Status Protocol & 0.01 & 0.09 \\
Data & 0.03 & 0.25 \\
Internet Control Message Protocol & 0.07 & 0.03 \\
Data & 0.06 & 0.00 \\
Address Resolution Protocol & 0.09 & 0.02 \\
\bottomrule
\end{tabularx}
\end{table*}

\end{document}